\documentclass[lettersize,journal]{IEEEtran}
\usepackage{amsmath,amsfonts}
\usepackage{algorithmic}
\usepackage{algorithm}
\usepackage{array}
\usepackage[caption=false,font=normalsize,labelfont=sf,textfont=sf]{subfig}
\usepackage{textcomp}
\usepackage{stfloats}
\usepackage{url}
\usepackage{verbatim}
\usepackage{graphicx}
\usepackage{cite}
\newtheorem{theorem}{Theorem}
\newtheorem{lemma}{Lemma}
\newtheorem{definition}{Definition}
\usepackage[justification=centering]{caption}
\usepackage{xcolor}

\begin{document}

\title{Achieving Maximum Efficiency in Schnorr-based Multi-signature and Applications in Blockchain} 

\author{Peng Zhang, Fa Ge, Yuhong Liu$^{\ast}$

\thanks{P. Zhang and F. Ge are with the College
of Electronics and Information Engineering, Shenzhen University, Shenzhen 518060, China (e-mail:  zhangp@szu.edu.cn; 2433582174@qq.com).}

\thanks{Y. Liu is with the Department of Computer Science and Engineering, Santa Clara University, Santa Clara 95053, USA (e-mail: yhliu@scu.edu).}}

\markboth{IEEE Transactions on Dependable and Secure Computing}%
{Zhang \MakeLowercase{\textit{et al.}}: Achieving Maximum Efficiency in Schnorr-based Multi-signature and Applications in Blockchain}


\maketitle

\begin{abstract}
Multi-signature aggregates signatures from multiple users on the same message into a joint signature, which is widely applied in blockchain to reduce the percentage of signatures in blocks and improve the throughput of transactions. The $k$-sum attacks are one of the major challenges to design secure multi-signature schemes. In this work, we address $k$-sum attacks from a novel angle by defining a Public Third Party (PTP), which is an automatic process that can be verifiable by the public and restricts the signing phase from continuing until receiving commitments from all signers. Further, a two-round multi-signature scheme MEMS with PTP is proposed, which is secure based on discrete logarithm assumption in the random oracle model. As each signer communicates directly with the PTP instead of other co-signers, the total amount of communications is significantly reduced. In addition, as PTP participates in the computation of the aggregation and signing algorithms, the computation cost left for each signer and verifier remains the same as the basis Schnorr signature. To the best of our knowledge, this is the maximum efficiency that a Schnorr-based multi-signature scheme can achieve. Further, MEMS is applied in blockchain platform, e.g., Fabric, to improve the transaction efficiency.
\end{abstract}


\section{Introduction}
\IEEEPARstart{M}{ulti-signature} allows multiple users to sign a message and aggregates all signatures into a joint signature, which can siginificantly reduce the communication overhead and improve the verification efficiency. As we know, digital signature is used extensively in blockchain to guarantee the integrity and non-repudiation of transactions. However, when multiple parties are involved in an approval process, verifying and storing individual signature from each party is costly. Multi-signature schemes, which can aggregate signatures from different parties into a joint signature with the same signature length, can play an important role in shrinking the size of block and extending blockchain ability.
For instance, Maxwell et al. \cite{maxwell2019simple} suggested using multi-signature to shrink the transaction data associated with Bitcoin Multisig addresses. Xiao et al.  \cite{xiao2020secure} deployed multi-signature in Fabric so as to reduce the size of transactions and improve the efficiency of endorsement and ledger update. Drijvers et al. \cite{drijvers2020pixel} proposed a forward-secure multi-signature scheme, which is applied to Proof-of-Stake Consensus in blockchain and yields notable savings in storage, bandwidth, and block verification time.

Based on the different difficulty assumptions and basis signature schemes, there are three major categories of multi-signature schemes, as RSA-based, BLS-based, and Schnorr-based multi-signature schemes. When one uses a 2048-bit modulus, the corresponding signature lengths for RSA, BLS, and Schnorr based schemes are 2048 bits, 224 bits, and 448 bits, respectively. Although BLS-based scheme achieves the smallest signature length, its high computational cost can not be overlooked. Due to their advantages in balancing the computational complexity and required storage space, Schnorr-based multi-signature schemes attract extensive research attentions recently. 

However, the earlier Schnorr-based multi-signature schemes \cite{Horster1995} are vulnerable to rogue-key attacks \cite{bellare2006multi}. In such attacks, an adversary chooses its public key as a function of that of honest users, and forges multi-signatures easily. To resist rogue-key attacks,  Bellare et al. \cite{bellare2006multi} first presented a multi-signature scheme BN in the plain public-key model. After that, another multi-signature scheme MuSig was proposed in \cite{maxwell2019simple}, where the joint signature could be verified exactly as a standard Schnorr signature due to key aggregation. The shortage of BN and MuSig is that three-round interactions are needed during the signing phase, increasing the communication overhead.
Bagherzandi et al. \cite{bagherzandi2008multisignatures} proposed an improved scheme BCJ and reduced the signing protocol rounds to two by using homomorphic trapdoor commitments, and Maxwell et al. \cite{cryptoeprint:2018:068} proposed another two-round MuSig. 
However, in 2019, Drijvers et al. \cite{drijvers2019security} showed that  none of the existing two-round multi-signature schemes, including BCJ \cite{bagherzandi2008multisignatures} and MuSig \cite{cryptoeprint:2018:068}, can be proved secure under standard assumptions for $k$-sum problem. Based on $k$-sum problem, if a signer can choose his/her commitment according to the other co-signers’ commitments, he/she can forge a signature passing the verification successfully.
 
Therefore, how to resist $k$-sum attacks becomes one focus of Schnorr-based multi-signature recently. Up to now, there are several schemes to resist $k$-sum attacks, including three-round schemes BN \cite{bellare2006multi},  MuSig \cite{maxwell2019simple} and two-round schemes mBCJ \cite{drijvers2019security}, MuSig2 \cite{nick2021musig2}. Their major working mechanisms are briefly described as follows. 
\begin{enumerate}
\item An additonal preliminary round interaction with commitment hash. This mechanism is adopted by BN and MuSig, where three-round interactions are required in the signing protocol. Compared with the later two-round schemes, a preliminary round is needed to prevent a signer from changing his/her commitment based on the others' commitments. Specifically, a signer $i$ submits a hash value $t_i$ of his/her commitment $R_i$ (i.e., $t_i=H(R_i)$), before he/she transmits this commitment to other co-signers. In this case, the commitment of each signer is independent.
\item Random commitment parameters. Adopted by mBCJ, this mechanism has its commitment parameters determined by the output of a random oracle applied to the message, thus parameters in sign query phase and sign forge phase are different. In this case, the forged signature based on $k$-sum problem cannot pass the verification.
\item Commitment vector. This method was first used in MuSig2, where a signer $i$ uses a commitment vector $(R_{i1},\cdots,R_{ik})(k \ge 2)$ instead of a single commitment $R_i$, and defines a linear combination $\hat R_i  = \prod\nolimits_{j = 1}^k {R_{_{ij} }^{b_j } } $ as his/her commitment, where coefficients 
$b_j$ are the hash of  vector $( \prod\nolimits_{i = 1}^n {R_{i1}} ,\cdots, \prod\nolimits_{i = 1}^n {R_{ik}} )$.
As the commitment of each signer is related to the commitment vector of all signers, it's impossible to generate commitment just according to other co-signers' commitment vectors.
\end{enumerate}

In general, in order to resist $k$-sum attacks, there are additional communication or computation costs involved, by preventing any signer from choosing his/her commitment based on that of other cosigners. In fact, as long as the last signer, who is the last one to submit the commitment, is always honest, $k$-sum attacks can be prevented. But this assumption that the last signer is honest is unreasonable. If we could find an honest party or program, which takes the final step of generating commitment, the security issue of $k$-sum attacks would be solved without additional cost. In the fields of cryptology and its applications, we often assume that there is a Trusted Third Party (TTP). But the assumption that everyone can trust the same party is too strong and typically leads to a centralized solution. Inspired by smart contracts on blockchains, which can be distributed on peer-to-peer networks, publicly verifiable, and executed automatically, we define a new concept, Public Third Party (PTP), as follows.

\begin{definition}[Public Third Party]
A Public Third Party (PTP) is an honest party or a piece of program whose actions or execution steps can be (1) fully open and verified by the public, and (2) automatically run without external interruptions. 
\end{definition} 

In this paper, we proposed to leverage a smart contract as a PTP to complete the last step of generating the joint commitment. Please note that, different from a TTP, which relies on a strong assumption of everyone's trust, a PTP earns its trust through its transparency, publicly verifiability and automatic execution.

In summary, we propose a new Schnorr-based multi-signature scheme,  by introducing PTP instead of additional communication or computation consumption to defend against $k$-sum attacks.
The proposed scheme can be completed in two rounds, and achieves almost the same computational complexity as the basis Schnorr signature. Therefore, we name it as MEMS (Maximum Efficient Multi-Signature). Our major contributions are listed as follows.
 
\begin{itemize}
\item Smart contract is leveraged as a Public Third Party (PTP) that contributes to resist $k$-sum attacks in the proposed scheme MEMS. After all signers send their own commitment $R_i$ to PTP, PTP generates a timestamp $t$. As PTP is public, the timestamp $t$ is difficult to be manipulated or estimated by an adversary. As joint commitment is generated based on $R_i$ and $t$, a malicious signer cannot generate a valid commitment with $k$-sum problem.

\item Security reduction of the proposed scheme MEMS is proved. Let the timestamp generated by PTP be $t$. By programming the random oracle $w=H(t)$ and computing $W=g^w$, the challenger can know the joint commitment $R=W^n\prod R_i$ before the adversary. Thus, our scheme MEMS is proved secure under discrete logarithm assumption in the random oracle model.

\item The proposed scheme MEMS reduces the amount of communication connections that need to be established. Although mBCJ and MuSig2 are also two-round, each signer needs to communicate with the other $n-1$ signers during the signing process involving $n$ signers. Therefore, they require a total of $n(n-1)$ communication connections per semi-round. While, MEMS introduces a PTP, and all signers only need to establish communication with the PTP, which only requires $n$ communication connections per semi-round, significantly less than that of other schemes. 

\item The proposed scheme MEMS can achieve a low computation cost. Because each signer only needs to know the group element corresponding to the timestamp, the computation cost of signing process can be reduced to $1$exp. For the computational cost of the verification, like most schemes, MEMS only needs 2 exponentiation operations. For each signer and verifier, the computation cost of MEMS is consistent with that of the basis Schnorr signature scheme.

We apply the proposed scheme MEMS in Fabric, one of the most popular blochchain platforms, for endorsement, and propose a modified Fabric transaction protocol mFabric, which can significantly improve the transaction efficiency. As MEMS aggregates multiple signatures into one joint signature, the proportion of signatures in a transaction gradually decreases. Meanwhile, the verification time for a block is shortened significantly.

\end{itemize}

\section{Related work}
\subsection{Schemes against Rogue Key Attacks}
Many existing studies consider the threat of rogue key attacks when
proposing multi-signature schemes. Since this kind of attacks is feasible both in theory and in practice,  existing multi-signature schemes take different measures to deal with this kind of attacks. 

Typically, there are two models to prevent rogue key attacks. One is \textit{Key Verification} (KV) model used in \cite{bagherzandi2008multisignatures,syta2016keeping,drijvers2019security}, where each public key is accompanied by a self-signed proof which is essentially a Schnorr signature.
The other is \textit{Plain Public-Key} (PPK) model proposed in BN \cite{bellare2006multi}, where the only requirement is that each signer must have a public key. This solution is to make each signer use a distinct \textit{challenge} $c_i$ when computing the partial signature
$s_i$. 
Ref. \cite{ma2010efficient} also adopted this idea, except that the \textit{challenge} $c_i$ is obtained by double hash. 
Later, Ref. \cite{maxwell2019simple,nick2021musig2,kilincc2021two} made a further improvement. They split \textit{challenge} $c_i$ into $a_i$ and $c$, so that the verification overhead is constant and independent of the number of signers.

Both methods have their pros and cons. For schemes adopting KV model, the aggregation of public keys is a simple product of all public keys. But each public key has a certificate, which is equivalent to increasing the length of the public key.
For schemes adopting PPK model, the overhead to compute aggregated public key is still linear to the number of signers.

\subsection{Schemes against $k$-sum Attacks}
In 2019, Drijvers et al. \cite{drijvers2019security} introduced sub-exponential attacks that apply to several multi-signature schemes \cite{bagherzandi2008multisignatures,ma2010efficient,syta2016keeping}. In such
attacks, adversary performing $k-1$ concurrent signing queries can create
a forgery in $\mathcal{O}(k\cdot2^{lg\ p/(1+lgk)})$ time and space, where $p$ is
the order of the group. 
In each signing query, the adversary chooses his/her commitment after the other co-signers. This results in that the hash on the product of 
commitments in $k-1$ queries equals to the sum of hashes on the commitments in query, based on the $k$-sum problem. Therefore, it is possible to forge a multi-signature on any message.

As mentioned earlier, how to defend against this kind of attacks becomes a focus recently.  
BN \cite{bellare2006multi} and MuSig \cite{maxwell2019simple} are immune to these attacks because they have three-round interactions in the signing phase. In the first preliminary round, each signer has to submit the hash on his/her chosen commitment, which forces each signer to determine its own commitment before learning others' commitments. 

Next, Drijvers et al. \cite{drijvers2019security} proved that none of the existing two-round multi-signature schemes could be proved secure under standard assumption, and presented a variant of the BCJ scheme call mBCJ. In mBCJ, commitment parameters rely on the message, and thus can be different in the sign query phase and the sign forge phase. Hence, mBCJ is the first two-round multi-signature scheme, which is proved to be secure based on Discrete Logarithm (DL) assumption.

Nick et al.\cite{10.1145/3372297.3417236} proposed a special multi-signature scheme MuSig-DN to resist this attack, which uses a pseudo-random function to generate random numbers deterministically. The adversary needs to select a suitable random number from a huge random number space to successfully carry out the attack. Therefore, MuSig-DN, which is the first Schnorr multi-signature scheme with deterministic signing, is immune to attacks.

Recently, based on a stronger One-More Discrete Logarithm (OMDL) assumption, Nick et al. \cite{nick2021musig2} presented a variant of the MuSig scheme called MuSig2, which is another secure two-round scheme. It uses commitment vectors instead of a single commitment, and each signer generates the commitment related to all other signers' commitment vectors, which can effectively prevent $k$-sum attacks. Applying a similar idea as MuSig2, a two-round multi-signature scheme via delinearized witness DWMS \cite{kilincc2021two} was proposed, which is secure in the Algebraic Group Model (AGM) and the Random Oracle Model (ROM) under the assumption of the difficulty of the one-more discrete logarithm problem and the 2-entwined sum problem.

\subsection{Other Multi-signature Schemes} 
In addition to Schnorr-based multi-signature schemes,
there are BLS-based and lattice-based multi-signature schemes. BLS-based multi-signature schemes only require a designated signer to collect the signatures of all signing participants, and do not require multiple interactions, which is very suitable for distributed application scenarios. Boneh et al. \cite{boneh2018compact} constructs a pairing-based multi-signature scheme by replacing the Schnorr signature in \cite{maxwell2019simple} by BLS signature, which is also suitable for Bitcoin. Drijvers et al. \cite{drijvers2020pixel} proposes a forward-secure BLS-based multi-signature, which is designed to deal with the problem of posterior corruptions in PoS Blockchain.
Recently, Pan et al. \cite{DBLP:conf/acisp/PanCCY22} gives the first multi-signature scheme for ECDSA.

Considering attacks based on quantum computing, some lattice-based multi-signature schemes are proposed. Kansal et al \cite{kansal2021efficient} proposed a lattice-based multi-signature scheme that supports public key aggregation. In \cite{peng2020new}, simultaneous multi-signature and sequential multi-signature schemes based on lattice are
proposed to resist the quantum attack using the difficulty of average-case Short Integer Solution (SIS) problem.

\section{Preliminaries}

\subsection{Notation and Difficulty Problem}

\textbf{Notation} For a non-empty set $S$, $s\leftarrow_r S$ means to select an element uniformly and randomly from the set and assigns it to $s$. We let $(\mathbb{G},p,g)$ denote the group description where $\mathbb{G}$ is a cyclic group
of order $p$ and $g$ is the generator of $\mathbb{G}$.  
Let $\lambda$ be the security parameter and bit length of $p$.
For a signer $i$, assume the pair of his/her public key and private key are denoted by $(X_i, x_i)$ where $X_i=g^{x_i}$. Assume there are $n$ signers participating in the multi-signatures.
We define three cryptographic hash functions $H_0:\mathbb{G}^n\times\mathbb{G}
\rightarrow \mathbb{Z}_p$, $H_1:\{0,1\}^*\rightarrow \mathbb{Z}_p$, $H_2:\mathbb{G}\times\mathbb{G}\times\{0,1\}^*\rightarrow \mathbb{Z}_p$.

\begin{definition}[Discrete Log Problem]
For a group description $(\mathbb{G},p,g)$, we define  \textbf{Adv}$_{\mathbb{G}}^{dl}$
of an adversary $\mathcal{A}$ as
\begin{displaymath}
  \textbf{Adv}_{\mathbb{G}}^{dl} = Pr[X=g^x:X\leftarrow_r \mathbb{G},x\leftarrow_r \mathcal{A}(X)]
\end{displaymath}
where the probability is taken over the random draw of $X$ and random coins of $\mathcal{A}$.
$\mathcal{A}$ $(t,\varepsilon)$-breaks the discrete log problem if it runs in
time at most $t$ with \textbf{Adv}$_{\mathbb{G}}^{dl}$ at least $\varepsilon$. 
Discrete log is $(t,\varepsilon)$-hard if no such adversary exists.
\end{definition}

\subsection{Generalized Forking Lemma}
Our security proof requires the Generalized Forking Lemma \cite{bellare2006multi} which extends Pointcheval and Stern’s Forking Lemma\cite{pointcheval2000security}. The proof of this lemma can be seen in \cite{bellare2006multi}.

\begin{lemma}\label{lemma1}
Fix an integer $q\ge 1$. Let $\mathcal{A}$ be an algorithm which takes as input \textbf{inp} and randomness $f=(h_1,\cdots,h_q,\rho)$, where $\rho$ is $\mathcal{A}$'s 
random coins and $h_1,\cdots,h_q$ are random values from $\mathbb{Z}_p$, and 
returns the failure symbol or a pair ($i$,\textbf{out}), where $
1 \le i \le q $ and \textbf{out} is the side output. The accepting 
probability of $\mathcal{A}$, denoted as \textbf{acc}($\mathcal{A}$), is defined as
the probability, over the random selection of \textbf{inp}, $h_1,\cdots,h_q\leftarrow_r
\mathbb{Z}_p$, and the random coins $\rho$ of $\mathcal{A}$, that $\mathcal{A}$
returns a non-$\perp$ output. The forking algorithm \textbf{Fork}$^{\mathcal{A}}$
associated to $\mathcal{A}$ is a randomized algorithm that takes input \textbf{inp}
as described in Algorithm \ref{algo1}. Let \textit{frk} be the probability that 
\textbf{Fork}$^\mathcal{A}$ returns a non-$\perp$ output. Then
\begin{displaymath}
  frk \ge \textbf{acc}(\mathcal{A}) (\frac{acc(\mathcal{A})}{q}-\frac{1}{p})
\end{displaymath}
\end{lemma}
\begin{algorithm}[H]
    \caption{\textbf{Fork}$^{\mathcal{A}}$(\textbf{inp})}\label{algo1}
    \begin{algorithmic}[1]
        \STATE{pick random coins $\rho$ for $\mathcal{A}$;}
        \STATE{$h_1,\cdots,h_q\leftarrow_r \mathbb{Z}_p$;}
        \STATE{$\alpha \leftarrow \mathcal{A}$(\textbf{inp}, $h_1,\cdots,h_q$, $\rho$);}
        \IF{$\alpha=\perp$}
            \STATE{return $\perp$;}
        \ELSE
            \STATE{parse $\alpha$ as ($i$, out);}
        \ENDIF
        \STATE{$h_i',\cdots,h_q'\leftarrow_r \mathbb{Z}_p$;}
        \STATE{$\alpha'\leftarrow \mathcal{A}$(\textbf{inp}, $h_1,\cdots,h_{i-1},h_i',\cdots,h_q',\rho$);}
        \IF{$\alpha'=\perp$}
            \STATE{return $\perp$;}
        \ELSE
            \STATE{parse $\alpha'$ as ($i'$, out$'$);}
        \ENDIF
        \IF{$i=i'$ and $h_i\neq h_i'$}
            \STATE{return ($i$, out, out$'$);}
        \ELSE
            \STATE{return $\perp$;}
        \ENDIF
    \end{algorithmic}
\end{algorithm}

\subsection{Rogue Key Attacks}
In rogue key attacks, the adversary is one of the involved signers denoted by $n$ who, for the vulnerability in the key setup phase, chooses ${x_n} \leftarrow_r \mathbb{Z}_p$ and uses a function of the honest signer's public key as its public key, i.e. 
\begin{equation*}
    X_n=(X_1X_2\cdots X_{n-1})^{-1}g^{x_n}
\end{equation*}
With this specially chosen public key, the adversary can arbitrarily 
forge the signature, since the aggregated secret key $x_n$
corresponding to the aggregated public key $\widetilde{X}$ is known to the adversary.
\begin{equation*}
    \widetilde{X}=\prod_{i=1}^nX_i=g^{x_n}
\end{equation*}

\subsection{$k$-sum Attacks}

We further elaborate the attack to multi-signature based on the $k$-sum problem \cite{drijvers2019security}. 
For simplicity, we consider the case where only two signers participate. The honest signer owns the keys $(X_1, x_1)$, and the adversary owns the keys $(X_2,x_2)$. 
The adversary concurrently opens $k-1$ signing oracle queries with the
honest signer on arbitrary message $m$. Let $j$ denote an index of the signing oracle query. According to \cite{drijvers2019security}, the adversary will perform the following steps:
\begin{enumerate}
  \item For each query $j\in [1,k-1]$, the adversary retrieves the public key $pk_1=g^{x_1}$ and a commitment $R_1^{(j)}=g^{r_1^{(j)}}$ from the honest signer. Then it creates $k$ empty lists $L_1,\cdots,L_k$ of equal size $s_L$, which will be filled with random elements next.

  \item For each list $L_j(j\in[1,k-1])$, the adversary picks $r_2^{(j)}\leftarrow_r \mathbb{Z}_p$, and computes $R_2^{(j)}=
  g^{r_2^{(j)}}$ as its commitment against $R_1^{(j)}$. It continues
  to compute $c^{(j)}=H_2(R_1^{(j)}\cdot R_2^{(j)},m,pk_1)$ and add
  $c^{(j)}$ to list $L_j$. The adversary repeats this process $s_L$ times for   each list $L_1,\cdots,L_{k-1}$.

  \item For list $L_k$, the adversary picks messages $m^*$ randomly, and adds $c^{(k)}=H_2(\prod_{j=1}^{k-1}R_1^{(j)},m^*,pk_1)$ to list $L_k$. It repeats this process until $L_k$ has $s_L$ elements.

  \item The adversary can get specific $c^{(j)}$ from list $L_j$ to solve $k$-sum problem in terms of Wagner's algorithm\cite{wagner2002generalized}, namely
  \begin{align*}
    c^{(1)}+\cdots+c^{(k-1)}\equiv c^{(k)}\ mod\ p \label{eq1}
  \end{align*}
  Next, it finds the $R_2^{(j)}$ as its commitment for all $j\in[1,k-1]$
  corresponding to $c^{(j)}$. Then, it responds $R_2^{(1)},\cdots,R_2^{(k-1)}$ to the honest signer, which are exactly corresponding to $k-1$ signing oracle queries.

  \item The adversary receives partial signatures $s_{1}^{(1)},\cdots,s_{1}^{(k-1)}$ from the honest signer where $s_{1}^{(j)}=r_1^{(j)}+c^{(j)}a_1x_1$. Now, all signing oracle queries come to the end. The adversary gets the message $m^*$ corresponding to $c^{(k)}$ (also called $c^*$) for which a forgery will be generated.
  
  \item Finally, adversary produces a forgery $\sigma=(c^*,s^*)$ where
  $s^*=\sum_{j=1}^{k-1}s_{1}^{(j)}+c^*a_2x_2$.

\end{enumerate}

So that is, based on the $k$-sum problem, the adversary performing $k-1$ concurrent signing queries can create a forgery in $\mathcal{O}(k\cdot2^{lg\ p/(1+lgk)})$ time and space, where $p$ is the order of the group.  

An important premise for this attack to be effective is that the adversary knows the $R_i$ of all other co-signers before submitting its own $R$. 
Therefore, in the proposed scheme, we address this attack by allowing the adversary to  compute $c$ only if it submits its $R$, which deters the adversary's behavior from step (2).

\section{A multi-signature scheme with public third party}
In order to defend against $k$-sum attacks, different from the existing schemes involving extra communication or computation costs, we propose to introduce a Public Third Party (PTP) (see \textit{definition 1}) to take the final step of generating commitment, which is the least cost way. The definition and construction of the proposed multi-signature with PTP are presented in this section. 


\subsection{The Definition}

The multi-signature scheme with the security parameter $\lambda$ consits of the following algorithms.
\begin{itemize}
    \item \textbf{ParamGen}($\lambda$) $\rightarrow par$: It generates the parameters of the signature scheme
    $par$ with respect to the security parameter $\lambda$.
    \item \textbf{KeyGen}$(par)\rightarrow (pk_i, sk_i)$: For any signer $i$, it generates a public/private key pair $(pk_i,sk_i)$ with the input $par$.
    \item \textbf{Agg}$(pk_1,\cdots,pk_n)\rightarrow pk_{agg}$: Input the public keys $pk_1,\cdots,pk_n$ of $n$ signers, and output the aggregated public key $pk_{agg}$.
    \item \textbf{Sign}$(par,pk_{agg},\{sk_1,\cdots,sk_n\},m)\rightarrow \sigma$: It is an interactive algorithm which runs between $n$ signers with private keys ${sk_1,\cdots,sk_n}$ and a PTP, in order to sign a common message $m\in \mathcal{M}$, where $\mathcal{M}$ is the message space. $\sigma$ is the signature output.
    \item \textbf{Verify}$(par,pk_{agg},m,\sigma)\rightarrow 1/0$: It verifies whether the signature $\sigma$
    is signed by the parties with the aggregated public key $pk_{agg}$ for the message $m$.
\end{itemize}

\subsection{The Proposed Scheme}
The specific multi-signature scheme is as follows:~\\

\noindent \textbf{Parameter Generation} (\textbf{ParamGen}($\lambda$)). Given security parameter $\lambda$, 
\textbf{ParamGen} generates a group $\mathbb{G}$ with prime $p$ order and a generator $g\in\mathbb{G}$. In the end, it outputs $par=(\mathbb{G},p,g)$.

\noindent \textbf{Key Generation} (\textbf{KeyGen}$(par)\rightarrow (pk_i, sk_i)$). Each signer $i$ randomly generates a secret private key $sk_i: x_i \leftarrow_r \mathbb{Z}_p$ and calculates the corresponding public key $pk_i: X_i=g^{x_i}$.

\noindent \textbf{Aggregation} (\textbf{Agg}$(pk_1,\cdots,pk_n)\rightarrow pk_{agg}$): Suppose $X_1,\cdots,X_n$ are the public keys of $n$ signers. For $i\in\{1,\cdots,n\}$, PTP calculates 
    \begin{equation*}
        a_i=H_0(PK,X_i)
    \end{equation*}
    where $PK=\{pk_1,\cdots,pk_n\}$. And then, compute the aggregated public key $pk_{agg}$:  $\widetilde{X}=\prod_{i=1}^n{X_i}^{a_i}$, which is open to the public. 

\begin{figure}[!t]
    \centering
    \includegraphics[width=\linewidth]{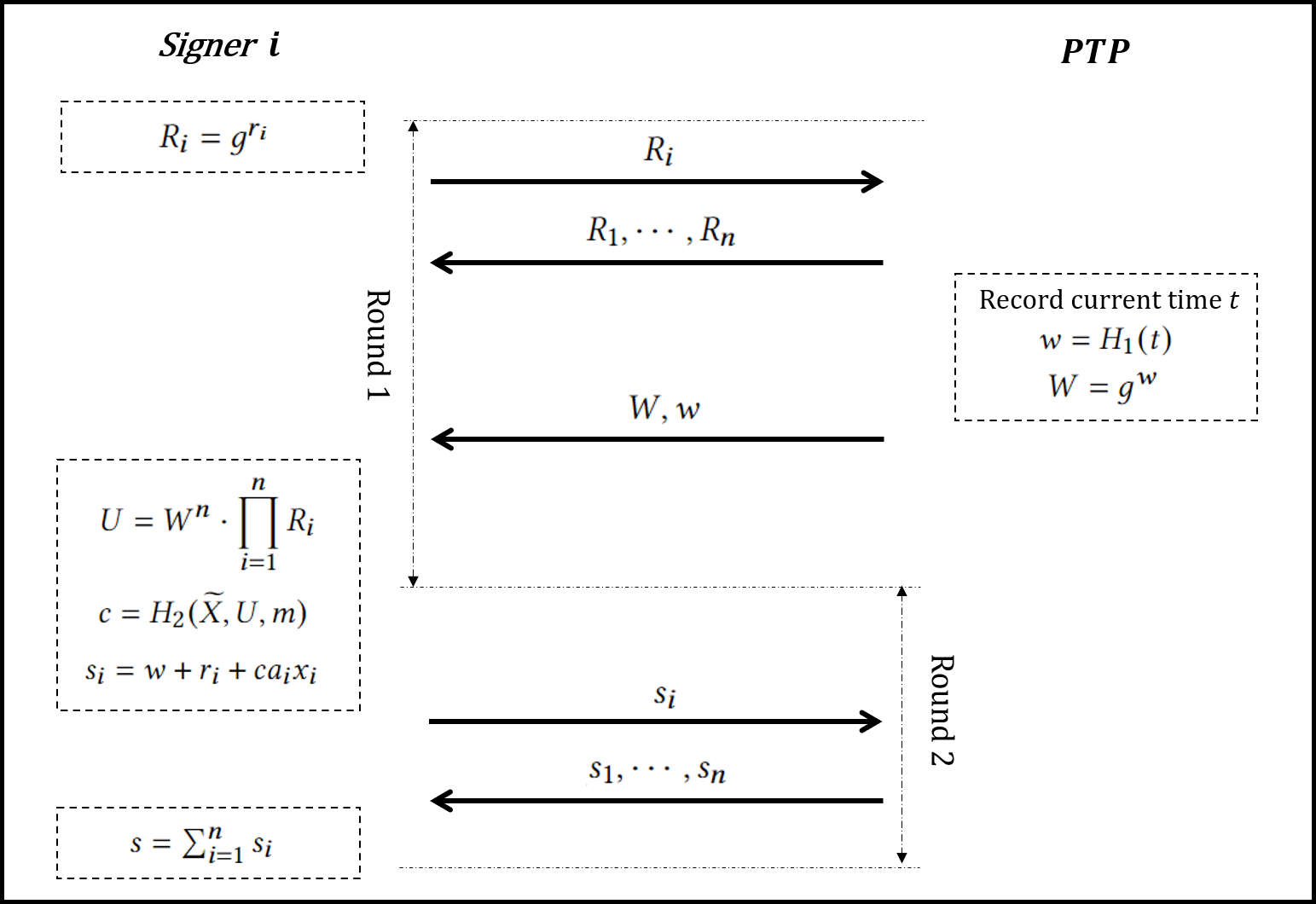}
    \caption{The Signing Process}
    \label{fig:1}
\end{figure}

\noindent \textbf{Signing} (\textbf{Sign}$(par,pk_{agg},\{sk_1,\cdots,sk_n\},m)\rightarrow \sigma$). As shown in Figure \ref{fig:1}, the signers will interact with a PTP in two rounds. 

\begin{itemize}
    \item \textit{\textbf{Round 1:}}\\
    First, the signer $i$ generates a random number $r_i\leftarrow_r \mathbb{Z}_p$, computes his/her commitment $R_i=g^{r_i}$ and sends $R_i$ to PTP. 
    
    Next, PTP waits to receive all commitments $\{R_i,\cdots,R_n\}$ from signers, and forwards them to all signers. 
    
    Last, PTP records the timestamp $t$ of current time, makes a hash $w=H_1(t)$ and computes $W=g^{w}$. PTP returns $w$ and $W$ to all signers.
    
    As PTP is automatic and public, the commitments $\{R_1,\cdots,R_n\}$ and timestamp $t$ are public and immutable.
    
    \item \textit{\textbf{Round 2:}}\\
    Upon reception of $\{R_1,\cdots,R_n\}$, $w$ and $W$ from PTP, any signer $i$ continues to compute the joint commitment
    \begin{align*}
        U&=W^n\cdot\prod_{i=1}^n{R_i},
     \end{align*}
     and the partial signature
     \begin{align*}
       s_i&=w+r_i+ca_ix_i \ (mod\ p),
    \end{align*}
    where $c=H_2(\widetilde{X},U,m)$, and send $s_i$ to PTP. 
    
    After collecting all partial signatures $\{s_1,\cdots,s_n\}$, PTP sends them to all signers.
    
    Finally, every signer can compute $s=\sum_{i=1}^ns_i\ mod\ p$ as the joint signature. 
    The output of this algorithm is $\sigma=(U,s)$.
\end{itemize}

\noindent \textbf{Verification} (\textbf{Verify}$(par,pk_{agg},m,\sigma)\rightarrow 1/0$). Given the aggregated public key $pk_{agg}=\widetilde{X}$, a message $m$ and a signature $\sigma=(U,s)$, 
the verifier calculates 
\begin{align*}
     c=H_2(\widetilde{X},U,m),
\end{align*}
and outputs $1$ if
\begin{align*}
    g^s=U\widetilde{X}^c.
\end{align*}

\section{Security proof}
\textbf{Proof Sketch}\quad With reference to \cite{maxwell2019simple}, we also use the forking lemma twice to prove security. Concretely, we first construct an algorithm $\mathcal{B}$
which runs the forger, simulating $H_0,H_1$ and $H_2$ uniformly at random and the
signature oracle by programming $H_2$, and returns a forgery together with extra
information about the forger execution. Later, we constructs an algorithm $\mathcal{C}$,
which runs the forking algorithm \textbf{Fork}$^\mathcal{B}$, obtaining two forgeries,
from which the discrete algorithm of aggregated public key can be extracted. Then, we
construct an algorithm $\mathcal{D}$, which runs the forking algorithm 
\textbf{Fork}$^\mathcal{C}$, solving the DL problem.

\subsection{Security Definition}
As there is a PTP except signers and  verifiers in multi-signature, this security model has made some modifications on the basis of \cite{bellare2006multi}. 
The security model requires the multi-signature to be unforgeable with the participation of at least one honest signer and a PTP, which cannot be corrupted by adversary. It's important to note that we assume that the PTP is publicly verifiable, which is different from the assumption that there is a trusted third party.
Without loss of generality, we assume that there is only one honest signer, and that the adversary corrupts all other signers (excluding PTP) and can choose their public keys
arbitrarily. Adversary can concurrently participate in any number of signature processes with an honest signer before returning a forged signature.

Next, we formally define the security game involving an adversary (forger) $\mathcal{F}$:
\begin{itemize}
    \item Randomly generate a pair of public key and secret key $(pk^*$, $sk^*)$ for the honest signer, using the public key $pk^*$ as the input for the forger $\mathcal{F}$.
    \item Adversary can participate in any number of signing processes with the honest signer. Formally, adversary can take as input a set of public keys $PK=\{pk_1,\cdots,pk_n\}$ where $pk^*$ occurs at least once and a message $m$ to access a sign oracle. This oracle implements the signature algorithm corresponding to the honest signer's $sk^*$, while the adversary plays the roles of the other corrupted signers in $PK$.
    \item At the end, the adversary has to output a set of public keys $PK=\{pk_1,\cdots,pk_n\}$, a message $m$, and a signature $\sigma$.
\end{itemize}

The adversary wins the game if $pk^*\in PK$, the forged signature is valid, and $\mathcal{F}$ never engages signature protocol for the public key set $PK$ and the message $m$.

\subsection{Security Analysis}

Our general idea is that assuming that there is a forger who can forge signatures for the above multi-signature scheme, by applying the forking lemma twice, the discrete logarithm problem can be solved with non-negligible probability. We use $(t,q_s,q_h,N,\varepsilon)$ to denote that the forger runs in time for up to $t$, makes at most $q_s$ signature queries, and at most $q_h$ hash queries, with a probability of success of at least $\varepsilon$, and has at most $N$ signers in the query process.

\begin{theorem}\label{the1}
Assume that there exists a ($t,q_s,q_h,N,\varepsilon$)-\textit{forger}
$\mathcal{F}$ against the multi-signature scheme with group parameter
$(\mathbb{G},p,g)$ and hash functions $H_0,H_1,H_2$ modeled as random oracles. Then, there exists an
algorithm $\mathcal{D}$ which $(t',\varepsilon')$-solves the DL 
problem for $(\mathbb{G},p,g)$, with $t'=4t+4Nt_{exp}+O(N(q_h+q_s+1))$
where $t_{exp}$ is the time of an exponentiation in $\mathbb{G}$ and 
\begin{align*}
  \varepsilon'\ \ge\  & \frac{\varepsilon^4}{(q_h+q_s+1)^3}\ -\ \frac{4q_s(q_s+q_h+1)}{p} \\ & -\ 
    \frac{12(q_s+q_h+1)^2+3}{p}.
\end{align*}
\end{theorem}

\begin{IEEEproof}
We construct an algorithm $\mathcal{B}$ which simulates random oracles $H_0$, $H_1$ and $H_2$ using three empty tables $T_0$, $T_1$ and $T_2$ respectively, and initializes three counters \textit{ctr0}, \textit{ctr1} and \textit{ctr2} as zero. Then, algorithm $\mathcal{B}$ picks random coins and runs the forger $\mathcal{F}$ on input the public key $X^*$, replying to hash queries and signing queries from the forger as follows.
\begin{itemize}
    \item  Hash query $H_0(PK,X)$: (Assume $X^*\in PK$ and $X\in PK$.) If $T_0(PK,X)$ has not been defined, then $\mathcal{B}$ increments
    \textit{ctr0} by one, randomly assigns $T_0(PK,X') \leftarrow_r \mathbb{Z}_p$ for all $X'\in PK\verb|\| \{X^*\}$ and assigns $T_0(PK,X^*)=h_{0,ctr0}$. 
    Then it returns $T_0(PK,X)$.
    
    \item  Hash query $H_1(t)$: If $T_1(t)$ is undefined, then $\mathcal{B}$ increases \textit{ctr1} and 
    assigns $T_1(t)\leftarrow w_{ctr1}$. Then, it returns $T_1(t)$.  
    
    \item  Hash query $H_2(\widetilde{X},U,m)$: If $T_2(\widetilde{X},U,m)$ has not been defined, then $\mathcal{B}$ increments \textit{ctr2} by one and assigns $T_2(\widetilde{X},U,m)=h_{1,ctr2}$. Then, it returns $T_2(\widetilde{X},U,m)$.
    
    \item  Signature query $(PK,m)$: If $X^*\notin PK$, then $\mathcal{B}$ returns $\perp$ to the forger $\mathcal{F}$. Otherwise, it parses $PK$ as $\{X_1=X^*,X_2,\cdots,X_n\}$. If $T_0(PK,X^*)$ is undefined, it makes an ``internal" query to $H_0(PK,X^*)$ which will define $T_0(PK,X_i)$ for 
    each $i\in \{1,\cdots$, $n\}$, sets $a_i=T_0(PK,X_i)$, and computes $\widetilde{X}=\prod_{i=1}^n{X_i}^{a_i}$. Then, $\mathcal{B}$ increments \textit{ctr2}, lets $c=h_{1,ctr2}$, $w=w_{ctr1+1}$, 
    draws $s_1\leftarrow_r \mathbb{Z}_p$, and computes $R_1=g^{s_1}{(X^*)}^{-a_1c}g^{-w}$ 
    where $cw_{ctr1+1}$ will be assigned to the value of next undefined query to $H_1(t)$ later.
    $\mathcal{B}$ sends $R_1$ to PTP.
    Next, $\mathcal{B}$ waits for $\{R_2,\cdots,Rn\}$ sent by PTP. 
    After $\mathcal{B}$ collects $\{R_1,\cdots,R_n\}$, it computes $U=g^{nw}\prod_{i=1}^nR_i$. If
    $T_2(\widetilde{X},U,m)$ has already been defined, $\mathcal{B}$ sets 
    \textbf{Bad$_1$}=\texttt{true} and return $\perp$. Otherwise it sets
    $T_2(\widetilde{X},U,m)\leftarrow c$.
    Note that PTP has made query to $H_1(t)$ but not yet sent $t$ and forger $\mathcal{F}$ 
    does not know $t$ at this time. Then, $\mathcal{B}$ waits to receive $t$ from PTP and compares $w$
    with $T_1(t)$. If $w$ is not equal to $T_1(t)$, $\mathcal{B}$ sets
    \textbf{Bad$_2$}=\texttt{true} and return $\perp$. Otherwise, $\mathcal{B}$
    sends $s_1$ to PTP, completing the signature query.
    
\end{itemize}

 If $\mathcal{F}$ gives a forged signature $(U,s)$ for a public keys multiset $PK$ where $X^*\in PK$ and a message $m$, $\mathcal{B}$ verifyies that this is a valid forgery, and stops if invalid. If the forgery is valid, $\mathcal{B}$ takes extra steps. Let $i_0$ be the value of \textit{ctr0} when $T_0(PK,X^*)=h_{0,ctr0}$ is assigned and $i_1$ be the value of \textit{ctr2} when $T_2(\widetilde{X},U,m)=h_{1,ctr2}$ is assigned. If the assignment $T_2(\widetilde{X},U,m)=h_{1,i_1}$ occurs before the assignment $T_0(PK,X^*)=h_{0,i_0}$, $\mathcal{B}$ sets \textbf{BadOrder}=\texttt{true} and returns $\perp$. If there exists another multiset of public keys $PK'$ such that the aggregated public key $\widetilde{X}'$ (corresponding to $PK'$) is equal to $\widetilde{X}$ (corresponding to $PK$), $\mathcal{B}$ sets \textbf{PKColl}=\texttt{true} and returns $\perp$. Otherwise, it outputs $(i_0,i_1,PK,U,s,\textbf{a})$, where vector \textbf{a} = ($a_1,\cdots,a_n$). By construction, $a_i=h_{0,i_0}$ for each $i$ such that $X_i=X^*$, and the validity of the forgery implies

 \begin{equation}
    g^s=U\prod_{i=1}^n{X_i}^{a_ih_{1,i_1}}.
\end{equation}

After analysis, the following conclusions can be drawn.

\begin{lemma}\label{lemma2}
Assume that there exists a ($t,q_s,q_h,N,\varepsilon$)-forger $\mathcal{F}$
in the random oracle model against our multi-signature scheme with
group parameter ($\mathbb{G},p,g$) and let $q=q_h+q_s+1$. Then, there
exists an algorithm $\mathcal{B}$ that takes as input a uniformly
random group element $X^*$ from $\mathbb{G}$ and uniformly random
scalars $w_1,\cdots,w_{q}$, $h_{0,1},\cdots,h_{0,q}$ and $h_{1,0},\cdots,h_{1,q}$, and
with successful probability at least
\begin{displaymath}
  \varepsilon - \frac{q_s(q_h+q_s+1)}{p} 
     - \frac{3(q_h+q_s+1)^2}{p}
\end{displaymath}
outputs $(i_0,i_1,PK,U,s,\textbf{a})$ where $i_0,i_1\in \{1,\cdots,q\}$, 
$PK=\{X_1,\cdots,X_n\}$ is a multiset of public keys such that
$X^*\in PK$, \textbf{a} = $(a_1,\cdots,a_n)$ is a tuple of scalars
such that $a_i=h_{0,i_0}$ for any $i$ such that $X_i=X^*$
\end{lemma}

Next, we use $\mathcal{B}$ to construct an algorithm $\mathcal{C}$, which will rewind $\mathcal{B}$. Algorithm $\mathcal{C}$ runs \textbf{Fork} $^\mathcal{B}$ with $\mathcal{B}$ as defined in Lemma \ref{lemma2}. At this time, \textbf{inp} refers to $X^*$, $w_1,\cdots,w_q$ and $h_{0,1},\cdots,h_{0,q}$. ($h_{1,1},\cdots,h_{1,q}$) play the role of ($h_1,\cdots,h_q$). $i_1$ plays the role of $i$. $(i_0,PK,U,s,\textbf{a})$ play the role of \textbf{out}.

Here we use the forking lemma for the first time to draw the following conclusions.

\begin{lemma}\label{lemma3}
Assume that there exists a ($t,q_s,q_h,N,\varepsilon$)-\textit{forger}
$\mathcal{F}$ in the random oracle model against the multi-signature
scheme with group parameters $(\mathbb{G},p,g)$ and let $q=q_h+q_s+1$.
Then, there exists an algorithm $\mathcal{C}$ that takes as input
a uniformly random group element $X^*\in \mathbb{G}$ and uniformly
random scalars $w_1,\cdots,w_q$, $h_{0,1},\cdots,h_{0,q}\in \mathbb{Z}_p$ 
and, with accepting probability at least
\begin{align*}
    \frac{\varepsilon^2}{q_h+q_s+1}-\frac{2q_sq+6q^2+1}{p}
\end{align*}
outputs a tuple ($i_0,PK,\textbf{a},\widetilde{x}$) where $i_0\in \{
1,\cdots,q\}$, $PK$ = $\{X_1,\cdots,X_n\}$ is a multiset of public keys
such that $X^*\in PK$, \textbf{a}=($a_1,\cdots,a_n$) is a tuple of 
scalars such that $a_i=h_{0,i_0}$ for any $i$ such that $X_i=X^*$,
and $\widetilde{x}$ is the discrete logarithm of $\widetilde{X}=
\prod_{i=1}^nX_i^{a_i}$ in base $g$.
\end{lemma}

We proceed to construct an algorithm $\mathcal{D}$ that rewinds algorithm $\mathcal{C}$. We let $q=q_h+q_s+1$. Algorithm $\mathcal{D}$ runs \textbf{Fork}$^\mathcal{C}$ with $\mathcal{C}$
as defined in Lemma \ref{lemma3}. At this time, $X^*$ and $w_1,\cdots,w_n$ play the role of \textbf{inp}. ($h_{0,1},\cdots,h_{0,q}$) play the role of ($h_1,\cdots,h_q$). $i_0$ plays the role of i. $(PK,\textbf{a},\widetilde{x})$ plays the role of \textbf{out}

Finally, we use the forking lemma for the second time, leading to Theorem \ref{the1}.
\end{IEEEproof}

\section{Performance analysis and application}
\subsection{Theorectical Analysis}
We compared the communication and computation cost of current two-round Schnorr-based multi-signature schemes, including mBCJ, MuSig2, DWMS, and our MEMS.
As shown in Table \ref{tab:spcom}, we analyze the size of the joint signatures, public key and secret key of a single signer, communication round and the total number of communications in the signing phase.
The proposed MEMS has the same and optimal signature size $|\mathbb{G}| + |\mathbb{Z}_p|$, public key size $|\mathbb{G}|$ and key size $|\mathbb{Z}_p|$ with MuSig2 and DWMS, and outperforms mBCJ. Although two-round interactions are needed, the total number of communications for MEMS is the lowest when $n \textgreater 3$, due to the introduction of PTP.

\begin{table*}[htb]
    \centering
    \renewcommand{\arraystretch}{1.3}
    \caption{Communication cost comparisons among two-round multi-signature schemes. 
    Here, $\mathbb{G}$ is the prime $p$ order group that these schemes work. $|\mathbb{G}|$ and $|\mathbb{Z}_p|$ denote the size of a group element and a scalar, respectively. 
    }
    \label{tab:spcom}
    \setlength{\tabcolsep}{6mm}
    \begin{tabular}{cccccc}
    \hline
     Scheme & Sig.size & pk.size & sk.size & Rounds & \#Comm  \\
     \hline
     mBCJ  \cite{drijvers2019security} & $2|\mathbb{G}|+3|\mathbb{Z}_p|$ & 
     $|\mathbb{G}|+2|\mathbb{Z}_p|$ & $|\mathbb{Z}_p|$ & 2 & $4n(n-1)$\\
    MuSig2 \cite{nick2021musig2} & $|\mathbb{G}| + |\mathbb{Z}_p|$ & $|\mathbb{G}|$
    & $|\mathbb{Z}_p|$ & 2 & $4n(n-1)$ \\
    DWMS \cite{kilincc2021two} & $|\mathbb{G}| + |\mathbb{Z}_p|$ & $|\mathbb{G}|$
    & $|\mathbb{Z}_p|$ & 2 & $4n(n-1)$ \\
    MEMS & $|\mathbb{G}| + |\mathbb{Z}_p|$ & $|\mathbb{G}|$
    & $|\mathbb{Z}_p|$ & 2 & 5$n$ \\
    \hline
    \end{tabular}
\end{table*}

In terms of computational cost, as shown in Table \ref{tab:comcos}, 
we analyze the consumption of \textbf{Sign}, \textbf{Verify} and \textbf{Agg} algorithms compared with mBCJ, MuSig2 and DWMS. Although public key aggregation happens in sign and verify algorithms for some schemes, we list it separately in Table \ref{tab:comcos} without loss of generality. As PPK model was used in MuSig2, DWMS and MEMS to resist rogue key attacks, \textbf{Agg} algorithm consumes $n$ exponentiation operations. Meanwhile, mBCJ uses KV model, and its \textbf{Agg} algorithm is implemented by multiplications,  so the time consumption is negligible. This is also the reason why the size of pubic keys of mBCJ is the largest.
The \textbf{Sign} algorithm only consumes $1$ exponentiation operation, which is one of the major advantages of the proposed MEMS scheme when compared with other multi-signature schemes. For the computational cost of verification, like most schemes, we only need 2 exponentiation operations. 
In terms of security, our scheme is resistant to the $k$-sum attacks, and a security proof is given in the ROM model based on the DL problem.

\begin{table*}[]
    \centering
    \renewcommand{\arraystretch}{1.3}
    \caption{ Computation cost and security comparisons among two-round multi-signature schemes with $n$ signers. $k$exp shows $k$-exponentiation in $\mathbb{G}$. Compared to exponential operations, the cost of addition, multiplication and hash is considered negligible. $v$ and $m$ represent the number of randomness in MuSig2 and DWMS, respectively. Typically, $v=4$ and $m=2$.}
    \label{tab:comcos}
    \setlength{\tabcolsep}{6mm}
    \begin{tabular}{cccccc}
    \hline
    Scheme & \textbf{Sign} & \textbf{verify} & \textbf{Agg} & $k$-sum & Security \\
    \hline
    mBCJ \cite{drijvers2019security} & 5exp & 6exp & 0 & Yes & DL, ROM\\

   MuSig2  \cite{nick2021musig2}& ($2v-1$)exp & 2exp & $n$exp & Yes & OMDL, ROM \\
   
    DWMS \cite{kilincc2021two}& $(mn+m)$exp & 2exp & $n$exp & Yes & OMDL, AGM,ROM\\
    
    MEMS & $1$exp & 2exp & $n$exp & Yes & DL, ROM \\
    \hline
    \end{tabular}
\end{table*}

\subsection{Experiment Analysis}
\textbf{\textit{Prototype:}} We implemented mBCJ, Musig2, DWMS and our MEMS in the Go Programming language. This experiment compares the time cost of \textbf{Sign} and \textbf{Verify} algorithms for each signer, not including aggregating public keys. Note that the implementation of DWMS does not organize signers into a spanning tree.

\textbf{\textit{Physical configuration:}} The experiments are performed in CentOS Linux
release 7.9.2009, with a Intel(R) Core(TM) i5-9500 CPU @ 3.00GHz processor and 4GB of 
RAM. In order to accurately measure the computational cost of the \textbf{Sign} algorithm, the round-trip delay of the network is ignored during the experiments.

\textbf{\textit{Experiment:}} We simulate the signing and verification process of the four
schemes to evaluate their computational overhead. Every experiment is repeated 10 times, and
the average of all the runs is compared.

\begin{figure}
    \centering
    \includegraphics[width=\linewidth]{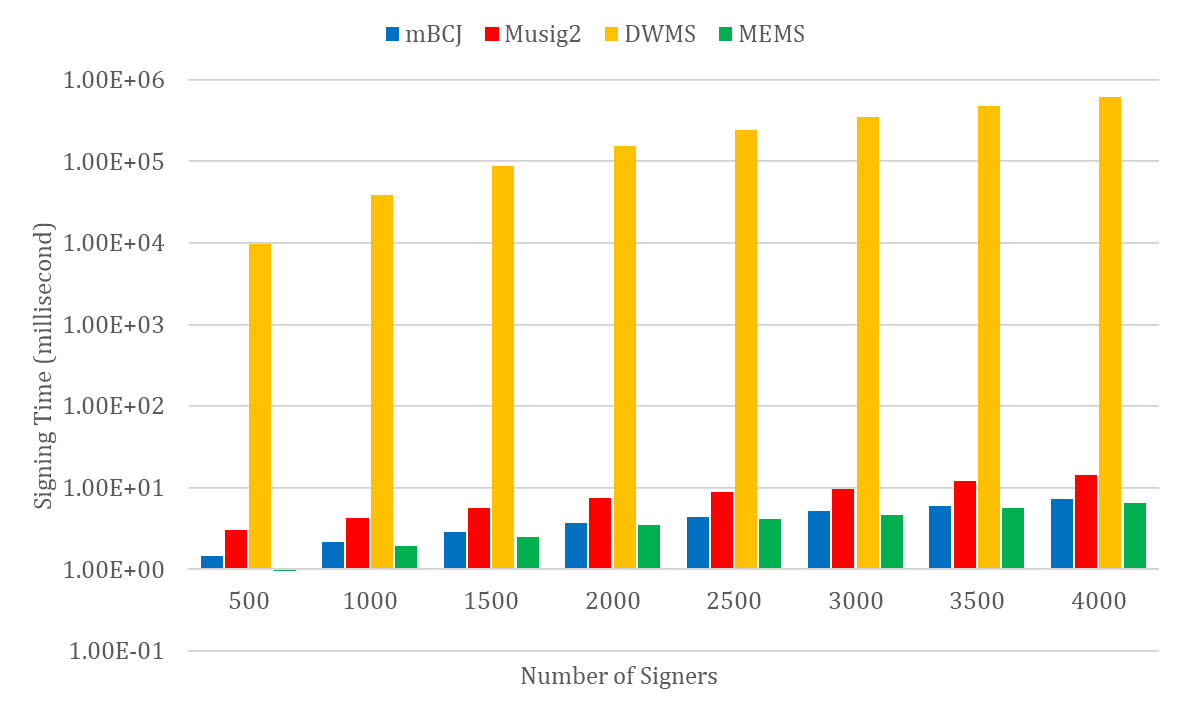}
    \caption{The time consumption for \textbf{Sign} algorithm ($y$-axis has logarithmic scale)}
    \label{fig:2}
\end{figure}

\begin{figure}
    \centering
    \includegraphics[width=\linewidth]{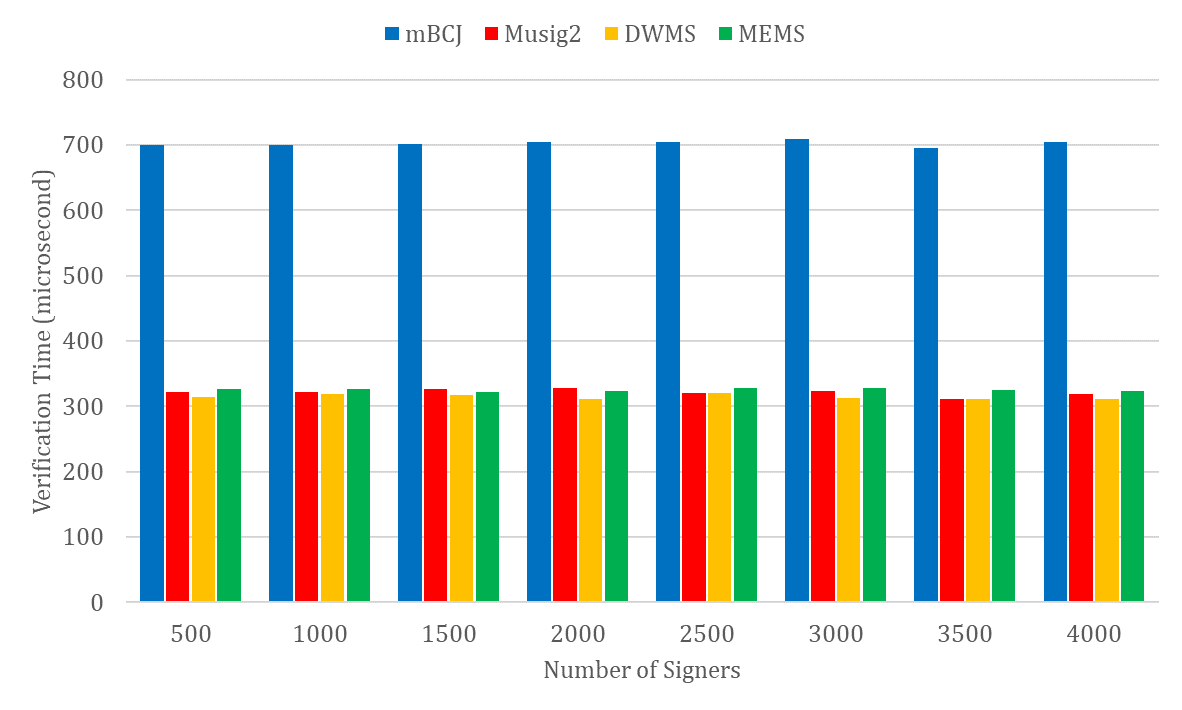}
    \caption{The time consumption for \textbf{Verify} algorithm}
    \label{fig:3}
\end{figure}

\textbf{\textit{Result:}} Figure \ref{fig:2} depicts the result of signing time, showing that one signer's signing time is no more than 10 milliseconds even if it scales to 4000 signers. 
Although the cost of an addition, multiplication, and hash is negligible compared to the exponent, a large number of these operations result in fluctuant running time.
Out of the four schemes, MEMS takes the least amount of time. 
On the contrary, DWMS increases the signing time rapidly as the number of signers increases. This is because the exponential operation required to calculate the joint commitment is linearly increasing when the number of signers increases.
Figure \ref{fig:3} shows that in our scheme, the verification time is consistent with most of the other schemes. Overall, the proposed  scheme achieves higher efficiency in the signature process, while maintaining the same verification efficiency as most of the other schemes. 

\section{Application in Blockchain Transactions}

Fabric is a permissioned blockchain platform with Certificate Authority (CA).
Whenever a client wants to create a new transaction or query the ledger, it needs to first connect to a peer node. 
Peer nodes are the fundamental elements of Fabric, responsible for managing ledgers and smart contracts.
Some peer nodes are called endorsers, which are responsible for simulating the execution of transactions and generating signature endorsements. 
One or more orderers provide the ordering service, and their main function is to pack transactions into a block and distribute it to peer nodes. 
Each peer node on the blockchain network, also called committing node, validates or invalidates the transactions in the blocks and commits valid blocks to its copy of the ledger. 

In order to ensure that the ledger remains consistent and up-to-date on all peer nodes in a channel, before a new transaction is committed to the ledger from a client, it must be endorsed by the peer nodes specified by the endorsement policy. Endorsement is the process where some endorsers simulate executing a transaction and return a proposal response to the client. The proposal response is the execution result of the smart contract on the peer nodes and contains a signature to prove that the node executed the smart contract. In most cases, the state update of the ledger requires the endorsement of the vast majority of peer nodes, which means that multiple signatures will be generated during the endorsement process, and the validity of the signatures needs to be verified multiple times. We apply our multi-signature scheme MEMS in the endorsement process to replace the original ECDSA signature, and optimize the endorsement process.

\begin{figure}
    \centering
    \includegraphics[width=\linewidth]{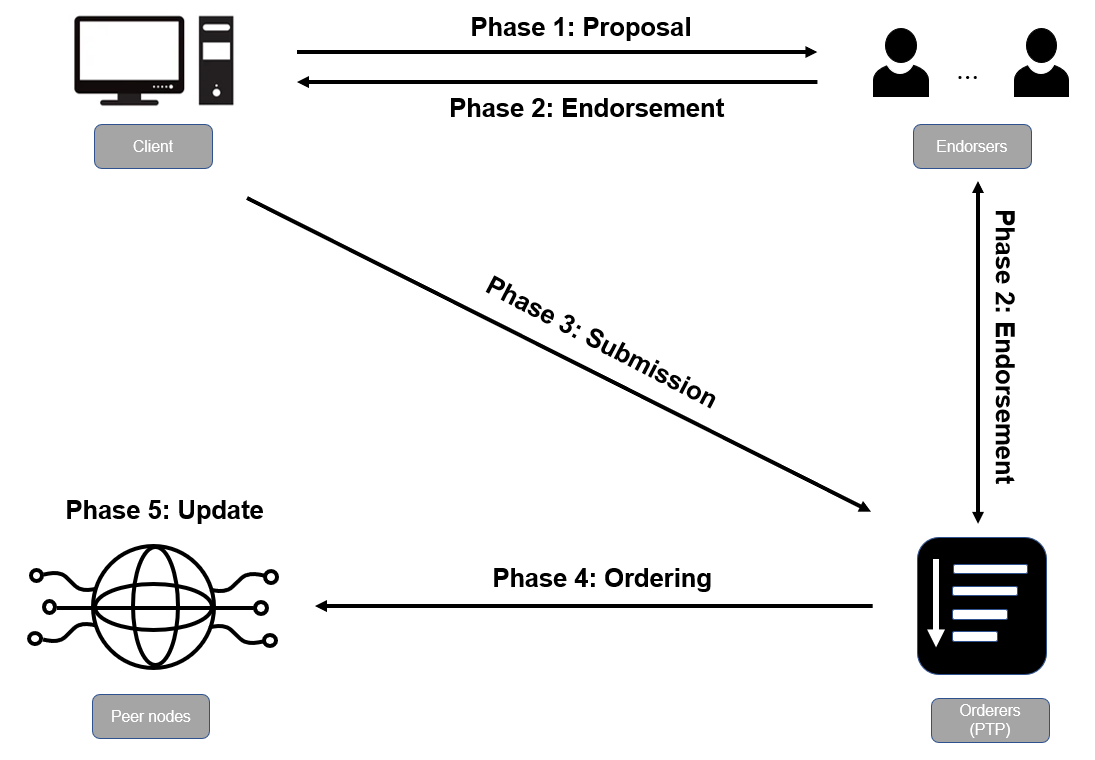}
    \caption{Our revised transaction process on Blockchain Fabric}
    \label{fig:mFabric}
\end{figure}

Before a transaction process begins, there are some preparations. First, the orderer needs to deploy a smart contract approved by the members in the network as the PTP. Second, Fabric CA uses \textbf{ParamGen}$(\lambda)$ to generate signature-related parameters $par$. After obtaining the public parameter $par$ from the CA, each peer node generates its own public and private key $(pk, sk)$ with \textbf{KeyGen}($par$). At this point, the preparation work has been completed. Next, we will introduce the revised transaction process, as shown in Figure \ref{fig:mFabric}.

\textit{Phase 1 (Proposal)}: The client generates a transaction proposal, signs the transaction proposal, and sends it to all endorsers specified by the endorsement policy.

\textit{Phase 2 (Endorsement)}: After the endorsers receive the transaction proposal, they first check whether the transaction proposal is accompanied by a valid signature of the client. Then, the endorsers execute the smart contract with the transaction proposal, generate a transaction proposal response, and call \textbf{Sign}($par, (pk, sk), m$) to sign the proposal response. In this process, the endorsers will interact with the smart contract on the orderer: the endorsers send the random number $R_i$ and the public key $pk_i$ to the smart contract, and the smart contract sends $(R_1,\cdots, R_n), (pk_1,\cdots, pk_n), w$, and $W$ to the endorsers. After that, all endorsers use $Agg(pk_1,\cdots, pk_n)$ to get $pk_{agg}$, proceed to the \textit{Round 2} of \textbf{Sign}. After an interaction with the orderer, each endorser obtains all partial signatures $s_1,\cdots, s_n$. Eventually, the endorser computes $s=s_1+\cdots+s_n$, appends it to the proposal response, and sends the proposal response to the client.

\textit{Phase 3 (Submission)}: After the client collects sufficient proposal responses from endorsers, it generates a transaction that contains the proposal and responses, and submits the transaction to the orderer.

\textit{Phase 4 (Ordering)}: The orderer receives transactions from different clients simultaneously, sorts these transactions in a predefined order, and packs the transactions into blocks, which are then distributed to the nodes in the network.

\textit{Phase 5 (Update)}: After the committing node receives a block, it first uses \textbf{Agg}$(pk_1,\cdots,pk_n)$ to obtain the aggregated public key $pk_{agg}$ of endorsers, and then calls the \textbf{Verify}$(par, pk_{agg}, m, \sigma)$ to verify the endorsement of each transaction in the block. If the signature is valid, the transaction satisfies the endorsement policy.

Next, we choose Fabric v2.2.8\footnote{https://github.com/hyperledger/fabric/tree/v2.2.8} to do some experiments. For the convenience of description, the transaction process on Fabric v2.2.8 is named Fabric, and the modified transaction process is referred to as mFabric. The computer configuration used in the experiment is the same as before.

\begin{figure}
    \centering
    \includegraphics[width=0.9\linewidth]{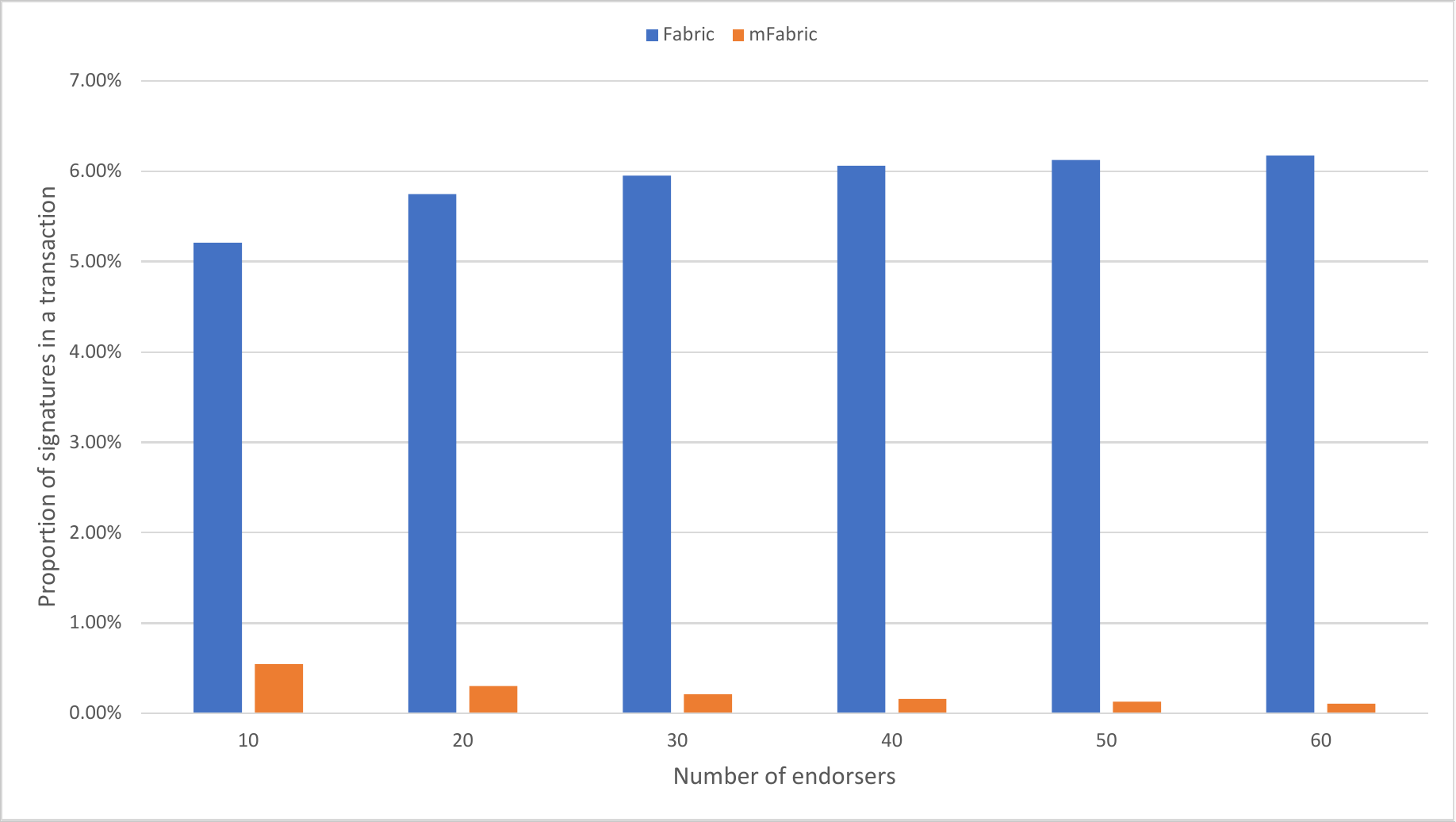}
    \caption{Proportion of signature in a transaction in Fabric and mFabric}
    \label{fig:5}
\end{figure}

First, we compare the proportion of signatures in each transaction of Fabric and mFabric. In Fabric v2.2.8, the structure of a transaction includes five parts, namely, header, signature, proposal, response and endorsement. The endorsement part contains the signatures of all endorsers generated during the endorsement process, and the number of signatures is the same as the number of endorsers. In mFabric, we modified the endorsement part to have only one signature no matter how many endorsers there are. The experimental results are shown in Figure \ref{fig:5}. It can be seen from the figure that as the number of endorsers continues to increase, the proportion of signatures in a transaction in Faric gradually increases, but the proportion of signatures in a transaction in mFaric gradually decreases, even approaching $0\%$.

\begin{figure}
    \centering
    \includegraphics[width=0.9\linewidth]{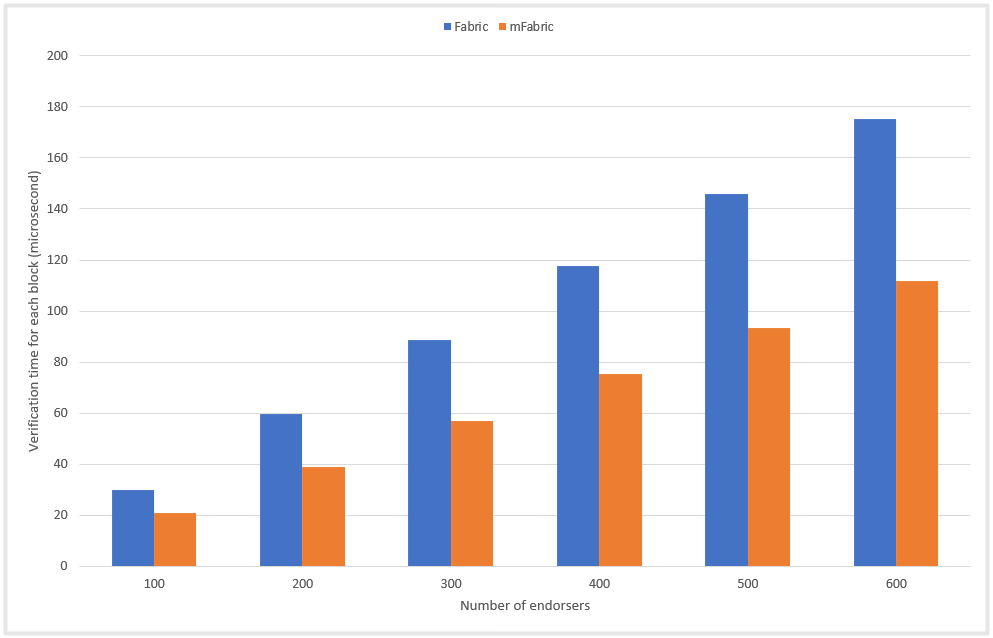}
    \caption{Time for a block containing only one transaction to be verified by a single node in Faric and mFaric respectively}
    \label{fig:6}
\end{figure}

Second, we compare the verification time on a single submitting node for a block containing only one transaction in Faric and mFabric. The experimental results are shown in Figure \ref{fig:6}. In both Faric and mFabric, as the number of endorsers gradually increases, the time for committing nodes to verify a single block will continue to increase. However, with the same number of endorsers, the verification time of a block in mFaric is always less than that in Fabric. This is because it is necessary to verify as many signatures as the number of endorsers in Fabric, while only one signature needs to be verified in mFabric.

Under the premise of keeping the signature time basically consistent, the comparison of the above two aspects shows that our multi-signature scheme cannot only reduce the space occupied by the signature in the block, but also improve the verification efficiency of the block. In fact, the modification of the network configuration in Fabric, including member changes, also requires the consent of multiple organizations, and it can also be improved with a multi-signature scheme. Therefore, multi-signature has a wide range of application scenarios in Fabric.

\section{Conclusion}

Inspired by Blockchain and the smart contract, a new method to resist $k$-sum attacks in Schnorr-based multi-signature is advanced through a public third party (PTP), which executes codes publicly and automatically  in a decentralized way. By requiring the recipient of all commitments as one execution condition of the smart contract, the proposed scheme can effectively prevent  $k$-sum attacks.
Based on this idea, a two-round multi-signature with maximum efficiency (MEMS) is proposed, which is proved to be secure based on discrete logarithm assumption in the random oracle model. Performance analysis shows that the proposed scheme MEMS achieves the lowest  communication and computation cost, compared to the other two-round Schnorr-based multi-signature schemes including mBCJ, MuSig2 and DWMS. As MEMS keeps the same efficiency as the basis Schnorr signature, we confirm that MEMS achieves maximum efficiency. Its applications in blockchain platform Fabric shows MEMS can reduce the space occupied by the signatures and improve the verification efficiency of the block.

\bibliographystyle{IEEEtran}
\bibliography{Submitted_version}

\end{document}